\documentclass{article}
\usepackage[a4paper,top=2cm,bottom=2cm,left=2cm,right=2cm,marginparwidth=1.75cm]{geometry}
\usepackage{hyperref}

\title{Activists Want Better, Safer Technology\footnote{This work originally appeared at the Workshop for Inclusive Privacy and Security (WIPS) 2020.}}
\author{Leah Namisa Rosenbloom \\ Brown University \\ leah\_rosenbloom@brown.edu}
\date{June 2020}

\begin{document}

\maketitle

\begin{abstract}
This work surveys 50 Black Lives Matter activists in the United States about the role of technology in organizing protests and other actions. Broad questions about the overall ease and safety of existing technology allowed them to spontaneously offer features and concerns that were most important to them. While the overwhelming majority (84\%) of activists reported using social media to find information, 64\% had concerns about data privacy or surveillance on social media, and 40\% had concerns about the credibility or reliability of information online. Community played a big role in activists' interactions with technology, with 72\% reporting that personal networks helped them find information or feel safe attending protests in general. We hope this study will provide a framework for deeper analysis in areas that are important to activists, so technologists can focus on building solutions that fully serve activists' needs and interests.
\end{abstract}
 
\section{Introduction}

Activists are a vulnerable population. They are frequently subjected to surveillance, censorship, and police targeting \cite{actsurv, cointelpro, actcen}. This is especially true for Black, Brown, and Indigenous movements, for example Black Lives Matter \cite{blmsurv} and the Indigenous Environmental Network \cite{stdrock}. 

Modern technology and in particular social media has made organizing for activism faster and easier than ever. Social media has already served as a cornerstone of many revolutions, for instance the Arab Spring \cite{arabspring}, Ukraine's Euromaidan Uprising \cite{euromaidan}, and Hong Kong's Umbrella Movement \cite{umbrella}. 

Modern technology and in particular social media has also made it easier than ever for governments and corporations to track and censor activists \cite{socmedsurv1, socmedsurv2, iransurv,corpsurv, syriasurv}. While there has been a significant amount of study on how activists use technology and combat surveillance \cite{socmedpro, safeorseen}, most work relies on digital evidence rather than in-person interviews or direct involvement. In addition, there has not yet been a push by computer scientists to study protest technology for the purpose of building better more suitable technology for activists.

The purpose of this study is to quantify using on on-the-ground interviews how activists are feeling about leveraging technology like social media to organize, and how they would improve the experience. This work asks: what about the technology makes it easy or difficult to use, and altogether desirable for organizing? What about the technology makes activists feel safe or unsafe? How could it be better? We hope to start a conversation between activists and technologists: what is it that activists really want from the technology they use, and how can we provide it?

\section{Methodology}

This work is a result of 50 interviews with Black Lives Matter activists in Philadelphia between June 1st and June 16th, 2020. Interviews were conducted in person directly following Black Lives Matter protests and other direct actions, and were limited to groups of one to four people. An essential distillation of interview content was recorded on paper at the time of the interview and later transcribed into digital format for coding and analysis. In order to preserve activists' privacy in a sensitive setting (heavy police and military presence), and in an effort to encourage open discourse, we did not take any personal information, voice, or video recordings of participants. We believe this setting and population is ideal for the study of activism, since it focuses on people who show up despite risks. 

\subsection{Interview Questions and Rationale}

The purpose of this work is to better understand the role that technology plays in organizing protests and direct actions, and how activists feel about the technology they use. We believe this understanding is necessary to inform the design of new technology that can fully serve activists' needs and interests. 

The following initial survey is meant to draw out the most important areas of concern for activists in a natural way. Questions are intentionally broad and up for interpretation; they ask activists to consider how ``easy'' it is to organize, and how ``safe'' they feel doing it. The role and reception of social media is of particular interest, though respondents also cited concerns with technology such as internet browsers and smartphones. Our hope is that future work will use these broader concerns as springboards for deeper analysis in each area.

\smallskip

\noindent \textbf{Question 1.} \textit{How did you hear about this action?}

\smallskip

\noindent \textbf{Question 2.} \textit{How do you usually find information about similar actions?}

\smallskip

\noindent \textbf{Question 3.} \textit{How easy is it to find the information you are looking for? Why do you think that is?} 
\smallskip

\noindent \textbf{Question 4.} \textit{How safe do you feel using social media to organize? Why do you think that is?}

\smallskip

\noindent \textbf{Question 5.} \textit{What might make digital organizing tools better (easier, safer)?}

\subsection{Coding for Key Factors}

The first two questions were easily coded using short keywords (`word of mouth', `Instagram', `email', etc.). Respondents could have any number of sources for either question. 

The ease and safety questions were separated into two parts: a scale from one to five (one being the not at all easy and not at all safe; five being very easy and very safe), and a rationale component that was coded at the granular level for specific word choices. Respondents' rationales were separated as much as possible at this level (for example, ``account verification'', ``organization verification'', and ``source verification'' are all separate codes) in case it is useful for future analysis. For the sake of this initial study, related codes are congealed together to form a broader picture (for example, the verification codes might be lumped together into a broader ``information authentication'' category). The improvement question was also coded at the granular level, then related codes were congealed for analysis. 

\section{Results}

The coded data for each respondent were entered into an array and parsed using Python. All of the raw data, as well as the Python structures and analysis, are freely available at \texttt{github.com/roguevillage}.

\subsection{Discovery}

We grouped discovery mechanisms into four categories: social media (Instagram, Facebook, and Twitter), other technology (YouTube, Reddit, email, text, websites, Google, and the news), physical (seeing the action, college campuses, or flyers), and personal networks (word of mouth, local groups, and work). 

While 42 out of 50 respondents (84.0\%) reported they used social media to find information on actions in general, only 24 out of 50 (48.0\%) reported that they heard of that day's action through social media. By contrast, 20 out of 50 respondents (40.0\%) tapped personal networks to find information in general, and 21 out of 50 (42.0\%) heard about that day's action through personal networks. We conclude that while the majority of activists use social media to find information about actions, traditional discovery methods (in particular word of mouth) have a more consistent yield of in-person participation.

The other two categories were far less popular. Only 11 out of 50 respondents (22.0\%) used other technology, and 4 out of 50 (8.0\%) used physical discovery to find out information about actions in general. 

The overall emphasis of discovery on social networks, both in person and online, is part of a larger theme of community engagement prevalent throughout the interviews. Further analysis of specific discovery methods and how they overlap is left for future work.

\subsection{Ease of Information Acquisition}
While coding the ease of use question, we realized we would need an exception to the traditional Likert scale: 13 out of 48 respondents (27.1\%) said that finding information was easy if a particular condition was met, and difficult otherwise. Of those respondents, 11 out of 13 (84.6\%) said that finding information was easy if and only if you were already connected, or ``in the know'' about particular groups, pages, and accounts. Similarly, 33 out of 50 respondents (66.0\%) cited existing connections as a determining factor in whether or not it was easy to find information online. 16 out of 50 (32.0\%) cited positive features of online platforms such as event pages and search functionality as contributing to ease of use. Overall, 7 out of 48 (14.6\%) found it very easy to find information, and 15 out of 48 (31.3\%) found it fairly easy.   

There were several factors contributing to difficulty in finding information. 10 out of 50 respondents (20.0\%) said they were hampered by the lack of credibility and reliability of the information on social media. Respondents described seeing events posted by unknown organizers, attending events that turned out to be different than advertised, and showing up to find events had been canceled last minute. They also reported discrepancies in the attribution of past events to various conflicting organizations (one respondent in particular cited dueling narratives about a group of counter-protesters---one side said they were white nationalists, the other side said they were protecting local store fronts). A second factor contributing to difficulty in finding information was timing: 6 out of 50 respondents (12.0\%) said that information dissemination mechanisms could not keep pace with current events. Counter to those who responded with praise for the organizational features of online platforms, 9 out of 50 respondents (18.0\%) reported problems with features, for example a lack of search functionality or feeling inundated with content, that made finding information difficult. Overall, 3 out of 48 respondents (6.3\%) found it very difficult to find information, and 10 out of 48 (20.8\%) found it fairly difficult. Correlations between easiness rationales, as well as between easiness rationales and other categories, is left for future work.

\subsection{Safety Concerns}

Respondents had a myriad of different reasons for feeling safe or unsafe using social media to organize. The biggest unsafe feeling, affecting 32 out of 50 respondents (64.0\%), was a general concern for the misuse of personal information, including privacy concerns (38.0\%), surveillance concerns (30.0\%), and data mining concerns (16.0\%). 11 out of 50 (22.0\%) cited concerns about the credibility and reliability of information, and 5 out of 50 (10.0\%) were concerned with hacking and device vulnerabilities. 9 out of 50 (18\%) expressed a general sentiment of fear and uncertainty surrounding the use of social media to organize. Other reasons given by one or two respondents included a lack of education (4.0\%) and a difference in values between the platform and the activists (2.0\%). 

In order to feel more safe, 7 out of 50 respondents (14.0\%) reported using privacy-enhancing technologies such as Signal, VPNs, private browsing, and firewalls. Some of those concerned with privacy on social media reported taking active steps such as identity hiding and self-censorship. 5 out of 50 (10.0\%) cited white privilege as creating a safe feeling, while 3 out of 50 (6.0\%) cited community. Other reasons given by one or two respondents included educational growth (2.0\%) and no prior thought (4.0\%). 

Overall, 6 out of 49 respondents (12.2\%) felt completely safe, whereas 12 out of 49 respondents (24.5\%) felt not at all safe. 16 out of 49 respondents (32.7\%) felt somewhat safe, and 10 out of 49 respondents (20.4\%) felt somewhat unsafe. The remaining 5 out of 49 (10.2\%) felt neither safe nor unsafe. These results suggest that while the activist population is split down the middle in terms of overall safety (44.9\% safe or somewhat safe; 44.9\% unsafe or somewhat unsafe), the feeling of total danger is more intense than the feeling of total safety. It is important to consider respondents' overall feelings on safety within the context of the rationales, specifically the majority's concern for privacy and security. Many respondents who felt safe still offered reasons for concern in the rationale, as evidenced by the relatively higher number of negative safety associations. 6 out of 50 (12.0\%) spontaneously offered that they only dealt with the risk of social media because there was no better alternative. Correlations between safety rationales, as well as between safety rationales and other categories, is left for future work.

\subsection{Improvement Recommendations}
Respondents had a lot of ideas about how to improve technology for organizing. First and foremost, 16 out of 48 respondents (33.3\%) expressed a desire to cut corporations out of the picture, starting with the removal of financial incentives and advertising (25.0\%), along with reduced power for social media companies and executives (12.5\%). Another big area for improvement was information authentication, with 10 out of 48 respondents (20.8\%) saying they would like to see an authentication mechanism like Twitter's blue check for trusted organizations and events. 8 out of 48 respondents (16.7\%) said they would like to see encryption on their organizing platforms, and 17 out of 48 (35.4\%) would like more privacy and less surveillance in general, including anonymity (8.3\%). 5 out of 48 respondents (10.4\%) supported some form of community control, for example platform decentralization or a community verification mechanism.

Other improvements given by fewer respondents were to increase user-friendliness (8.3\%) and security against hacking (8.3\%), educate people in digital rights and literacy (6.25\%), and establish value alignment between platforms and activists (2.1\%). Suggestions for new platform features offered by 7 of 48 respondents (14.6\%) included establishing protected categories for activists, a live feed option for real-time updates, a section in event pages for protest route information, and a consolidated event calendar for inter-organizational coordination. Correlations between improvement recommendations, as well as between improvement recommendations and other categories, is left for future work.

\subsection{Common Threads}

The biggest commonality across all sections was community: a total of 36 out of 50 (72\%) of respondents mentioned their personal networks as positive contributors to ease, safety, and/or improvement of digital platforms. Conversely, not having community was a big contributor to difficulty in finding information on actions. Next, a total of 32 out of 50 respondents (64.0\%) expressed some concern for the lack of privacy online, stemming primarily from data collection (10.4\%) and surveillance (34.0\%). Finally, 20 out of 50 respondents (40.0\%) expressed some concern for the lack of credibility and reliability of information they found online. 

\subsection{Questions for Consideration}

\noindent \textbf{How might we make it easier to build up social networks, while still maintaining privacy? How might we build authentication mechanisms that preserve privacy?} Social networks, authentication, and privacy might seem at odds with one another on the surface, but there are existing tools like anonymous credentials \cite{anoncred} that can connect and authenticate users without revealing their identities.

\smallskip

\noindent \textbf{How might we integrate features activists like about community strength into organizing platforms?} Social networks are popular because they allow users to form their own online communities. There are some new community-oriented features, such as community verification, inter-organizational event calendars, and values-based moderation that activists might like to see integrated into organizing platforms. 

\smallskip

\noindent \textbf{How might we replace corporate social media with better alternatives?} One respondent pointed out that while they would like to get rid of social media, new platforms generally do not have the pull and popularity necessary to transition the masses away from Facebook, Instagram, and Twitter. Because social networks require volume to function properly, it will be necessary for aspiring creators to consider the transitional phase. 

\section{Conclusion}

Good technology is informed by the needs of the population it is meant to serve. The technology we currently use, however, is not meant to serve us; it is meant to serve the financial interests of board executives and company stakeholders. Corporations will not forsake profit to protect vulnerable populations from surveillance and police targeting, fake news, and censorship. We hope this work will form a basis for the study of better technology for organizing, and that one day, vulnerable populations will be able to leverage technology to demand justice, without fear of retalliation.

\section{Acknowledgements}
Thank you to Bruno Fleischman Mayro for his help coding, and for being brave with me at protests. Thank you to Gigi, Annie, and Matt for being my first interview participants. Thank you to all of the activists who participated in this study, and to all those who continue to persevere despite the risks. Black Lives Matter.

\bibliography{bib}
\bibliographystyle{plain}

\end{document}